# Reversible Adversarial Examples with Beam Search Attack and Grayscale Invariance

Haodong Zhang, Chi Man Pun*, and Xia Du

Department of Computer and Information Science
University of Macau, Macau, China
* Corresponding author. Email: `cmpun@umac.mo`

**Abstract.** Reversible adversarial examples (RAE) combine adversarial attacks and reversible data hiding technology on a single image to prevent illegal access. Most RAE studies focus on achieving white-box attacks. In this paper, we propose a novel framework to generate reversible adversarial examples, which combines a novel beam search based black-box attack and reversible data hiding with grayscale invariance (RDH-GI). This RAE uses beam search to evaluate the adversarial gain of historical perturbations and guide adversarial perturbations. After the adversarial examples are generated, the framework RDH-GI embeds the secret data that can be recovered losslessly. Experimental results show that our method can achieve an average Peak Signal-to-Noise Ratio (PSNR) of at least 40dB compared to source images with limited query budgets. Our method can also achieve a targeted black-box reversible adversarial attack for the first time.

## 1 Introduction

Deep neural networks (DNNs) have made significant progress in various applications. However, studies have shown that well-designed adversarial attacks can easily fool a variety of existing deep learning models in multiple fields [10, 21, 22, 2] and have been an important issue in AI and multimedia security areas [3, 31, 28]. Adversarial examples are made by adding small perturbations to benign inputs. Humans can't perceive such differences, which can cause the network to make wrong predictions.

Reversible adversarial attacks are a special type of adversarial attack. It refers to embedding the information needed to recover secret data (such as the source image) into the adversarial example through reversible data hiding (RDH) technology [7, 19, 30]. Reversible adversarial examples (RAE) are generated by such attacks. It can confuse unauthorized classifiers from illegal access to images and becomes a carrier to save data and restore them when needed. RAE provides a significant framework for data privacy protection. Recent RAE studies mainly focus on recovering the source image from RAE losslessly. These works combine source images with adversarial attacks through existing RDH technology. Research [19] has proven RAEs preserve the privacy of image data.

Although the existing RAE framework allows users to keep data private in public spaces. The visual quality and attack capabilities of these examples still need to be improved. For example, many RAEs can only perform white-box attacks [30]. In this paper, we propose reversible adversarial examples with grayscale invariance. This framework combines beam search attack, a novel adversarial attack scheme to perform targeted black-box attacks and reversible data hiding with grayscale invariance (RDH-GI) technology to maintain the visual quality of colour images by maintaining grayscale. The beam search attack randomly generates adversarial noise at different timestamps and queries its output score, then use beam search to select those noises with the most significant adversarial gain. According to the direction of these noises, a new noise is generated by a decayed step size without query. This new noise and all randomly generated noises are added to the input image accumulatively to obtain an adversarial example. We also use RDH-GI [9, 13] to ensure image quality. RDH-GI perform data embedding on colour channels red and blue, while the green channel is used for grayscale adjustments to remove modifications from the red and blue channels. Our main contributions are summarized as follows:

– We proposed a novel beam search attack for generating reversible adversarial examples. It periodically makes a decision based on historical information of perturbation to reduce the number of queries with negligible additional cost.
– We are the first to propose the targeted black-box attack for reversible adversarial examples with grayscale invariance to improve the visual quality.
– Extensive experiments are performed to evaluate our method on various real-world datasets. Compared with other state-of-the-art methods, our method can achieve an attack success rate of nearly 90% in the black box setting.

The rest of this paper is organized as follows: We summarize the research related to the reversible adversarial attacks in the Sect. 2. In Sect. 3 we elaborate on the proposed framework. In Sect. 4, we conducted an extensive evaluation of experimental results. Conclusions are drawn in Sect. 5.

## 2   RELATED WORK

In this section, we investigate the adversarial attack methods and the reversible data hiding methods, which are the components of the RAE framework. We also introduce the latest research on reversible adversarial attacks.

### 2.1   Adversarial Attacks

We discussed adversarial attacks in terms of gradient-free Attacks and gradient estimation attacks. Gradient estimation attacks first estimate the gradients of the target model and then use them to run the attack. Many current RAE frameworks [19, 30] are based on such attacks. AutoZOOM [25] using gradient estimation based on random vectors significantly improves the performance. HopSkipJumpAttack [6] estimates the gradient direction using binary information at the decision boundary. QEBA [17] estimates the gradient through Monte Carlo simulation in a low-dimensional subspace and achieves a much lower magnitude of perturbation compared to existing decision-based attacks.

Gradient-free attacks do not attempt to estimate gradients but instead generate heuristic adversarial examples based on query results. Alzantot et al. [1] proposed GenAttack, which improves the genetic algorithm by performing dimensionality reduction of the search space and adaptive parameter scaling. Guo et al. proposed simple black-box adversarial attacks (SimBA) [11]. SimBA randomly samples a vector from a predefined orthonormal basis and either adds or subtracts it from the target image. Our framework implements the gradient-free attack through a score-based heuristic algorithm.

### 2.2 Reversible Data Hiding

Reversible Data Hiding (RDH) has broad application prospects in multimedia security. It can extract embedded hidden data from labelled camouflage images to achieve privacy. The RDH algorithm takes advantage of the empirical fact that the colour histograms of natural images are not uniformly distributed. According to this, additional information can be encoded into the image without being noticed by shifting the bins of the colour histogram.

Traditional RDH will distort the host image. Therefore, various RDH methods which aim to improve image quality are proposed. However, many RDH algorithms are presented for grey images [32], which are not as good as coloured images in terms of visual quality. Besides, many feature analyses of images are based on the grayscale domain, so maintaining grayscale invariance plays a significant role in practical applications. RDH with grayscale invariance (RDH-GI) [13], are designed for colour images. RDH-GI explores the relationship between R, G, and B channels. This method keeps the grayscale of the image unchanged. Specifically, it uses the R and B channels of the colour image to embed information and ensure grayscale invariance by adjusting the pixel value of the G channel. We adopt the RDH-GI technology to compose our reversible adversarial examples.

### 2.3 Reversible Adversarial Attack

The purpose of the reversible adversarial attack is to prevent unauthorized access to secret image data. Reversible adversarial examples (RAE) are a special type of adversarial attack. It refers to embedding the information needed to recover secret data (such as the original image of the adversarial example) into the adversarial example through reversible data hiding technology [18, 19, 29, 30]. Recent RAE studies tend to innovate in certification and data capacity. Liu et al. first proposed the construction of reversible adversarial examples [18], which are obtained by combining histogram modification-based reversible data hiding (RDH) and white-box attack. Yin et al. take advantage of Reversible Image Transformation (RIT) to construct reversible adversarial examples [30]. They also explored the reversibility of adversarial attacks under locally visible adversarial perturbation. The mentioned works mainly aim to combine source images with adversarial attacks through RDH technology, and tend to choose white box attacks as part of the framework (e.g., FGSM [10], C&W [5]). We will compare our work with these methods in Sect. 4.2.

## 3 Reversible Adversarial Attack with Grayscale Invariance

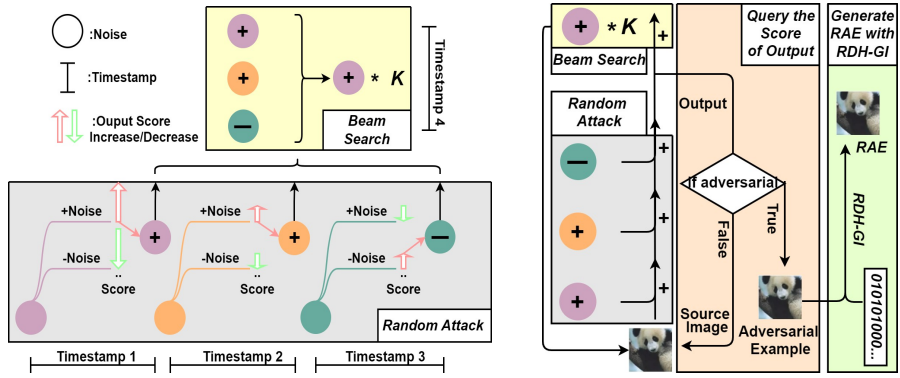

Fig. 1: Overvall framework of proposed reversible adversarial attack.

### 3.1 Preliminaries

In a black-box setting, the goal of our reversible adversarial attack is to find a reversible adversarial example $x'_{adv}$ with targeted class $y$ while minimising the accumulated perturbations, which can be formulated as equation (1).

$$\arg\max_{\delta+d} p(y|x'_{adv}),\ s.t.\ \min ||\delta+d||_2 \tag{1}$$

The distortion caused by adversarial perturbation and RDH represents as $\delta$ and $d$ in Eq (1). The symbol $||.||_2$ represents the $l2$ norm. In the RAE-GI framework, $\delta$ is mainly related to the query count, and $d$ is only influenced by the length of embedded message $I$ due to the existence of error correction bits. We write about the progress of generating RAEs $x'_{adv}$ in Eq (2).

$$x'_{adv} = RDH(x_{adv}, I) = RDH(x + \delta, I) \tag{2}$$

## 3.2 Overview

We propose a framework to generate reversible adversarial examples with grayscale invariance (RAE-GI). Figure 1 demonstrates the framework of our reversible adversarial attack, which mainly consists of the beam search attack and RDH-GI embedding. The key parts of the beam search attack are shown in detail. The query part of the random attack is simplified, while every step of a random attack needs a single query. The beam search attack mainly includes random attack and beam search. In the random attack part (timestamp 1, 2, 3), adding or subtracting randomly generated noise may increase the $p(y|x)$ of the source image. We query and record the noise that increases the score to the beam recorder set. In the beam search part (timestamp 4), we choose the noise direction with the largest score gain rate and generate new noise with decayed step size in this direction without query. After each timestamp, the resulting noise is added to the source image until an adversarial example is formed.

The RDH-GI technique generates the grey version of the adversarial image and modifies channels R and B to embed messages and generate RAE. The embedded messages can only be extracted when receivers have full knowledge of the RDH-GI progress.

## 3.3 Beam Search Attack

Reversible adversarial examples are one kind of adversarial example. Its basic framework is consistent with it. In targeted black-box settings, with a source input $x$, a target y, and accumulated adversarial perturbations $\delta$, attackers try to perturb $x$ to form the adversarial output $(x + \delta)$, maximizing $p(y|x + \delta)$ while $\delta$ should be minimized. The output probability plays an important role when we try to perform the query-based attack.

We demonstrate the pseudocode of the beam search attack in algorithm 1. Given a benign input $x$ and randomly generated same-size noise $q$, performing $(x+q)$ may increase the probabilities of $p(y|x+q)$. The beam search attack generates random noise based on the above ideas. When adding noise does not increase $p(y|x)$, the beam search attack will subtract this noise and query $p(y|x-q)$. This is similar to how BoundaryAttack [4], and simple black-box attack (SimBA) [11] update the adversarial noise. However, they do not reuse the direction $q$ generated by each generation, so whenever random noise is generated, the attacker needs to query $p(y|x)$ to decide whether to add this perturbation.

Inspired by [26], which uses the geometric properties to approximate the decision boundary, we exploit the direction of historic perturbations to reduce the number of queries using beam search. Assuming $x_i$ represents the noise generated at timestamp $i$, the beam size is $n$. We periodically record the ratio gain of $p(y|x_i)$ caused by the noise with direction $q$ in beam recorder $D$ when the timestamp count is within $n$. Then we perform beam search on set $D$ and select the direction with the largest gain to add perturbation without query. After that, we clear the beam recorder and enter the next loop. We can assume that each randomly generated noise $q$ has an approximately uniform contribution to the final cumulative noise $\delta$, so this beam search strategy can significantly reduce the number of iterations. It can also omit some steps of generating new random noise.

---

**Algorithm 1:** Baseline of Beam Search Attack

**Input:** source $x$, target $y$, score $p$, step size $\alpha$, beam recorder $D$, beam size $n$, decay coefficient $k$
**Output:** adversarial example $x + \delta$
$\delta \leftarrow$ initialization();
**while** *iteration count <limit* **do**
    Randomly generate direction $q$;
    $p_{in} = p(y \mid x + \delta)$;
    $p_\delta = p(y \mid x + \delta + \alpha q)$;
    **if** $p_\delta > p_{in}$ **then**
        $\delta = \delta + \alpha q$;
        Record $D: q \leftarrow p_\delta/p_{in}$;
    **else**
        $\delta = \delta - \alpha q$;
        $p'_\delta = p(y \mid x + \delta - \alpha q)$;
        Record $D: -q \leftarrow p'_\delta/p_{in}$;
    **end**
    **if** *iteration count % n = 0* **then**
        $\delta = \delta + k\alpha q[\mathbf{max}(p_\delta/p_{in}) \text{ in } D]$;
        $D$.clear()
    **end**
**end**

---

**Decayed Learning Rate.** Assuming any directions may increase $p(y|x)$. If we directly perform a beam search on the direction set with the initial learning rate (or step size), it will cause the results of the beam search to determine the direction of the perturbation change to a certain extent. For example, we set the beam size to 3, which means beam search performs an alternative attack every 4 queries. We suppose the direction made by each query has a similar contribution to the final result. Thus, the direction of the accumulated adversarial perturbation may have 1/4 part from the beam search, which may not be the desired result. To solve this problem, we choose to proportionally reduce the learning rate (by default 1/3) when conducting beam search attacks, which can use historical direction information without unduly affecting the direction of accumulated perturbation. In addition, when the output probability increases to a certain extent, the randomly generated adversarial perturbation direction may no longer improve the attack result, which means that the return value of the beam search is 0. In this case, we do not perform a beam search attack, similar to the case of BoundaryAttack [4]. Empirically, the above situation is difficult to occur in low-frequency attacks [11], so we do not need to introduce additional calculations.

**Low Frequency Subspace.** In general, the low-frequency subspace of the image records the most vital information, so random noise in the low-frequency space is more likely to be adversarial [11, 17]. Thus we use discrete cosine transformation to reduce the search space. Discrete cosine transform (DCT) represents an image as a sum of sinusoids at different frequencies. In our proposed method of beam search attack, adversarial perturbations generated in the random attack part are sampled from the orthonormal low-frequency subspace via discrete cosine transformation.

### 3.4 Reversible Data Hiding with Gray Invariance

RAE can utilize the RDH scheme to embed additional information into the adversarial image in order to make the receiver output the desired result. Recent studies of RAEs mainly focus on recovering the source image from the RAE [19, 30]. Considering that adversarial example does not affect human understanding of images, we use RDH-GI (reversible data hiding with grayscale invariance) to embed general information in this paper.

The grayscale of the image can be regarded as the combination of the grayscale of the R, G, and B channels. Assuming that $r$, $g$, and $b$ represent the intensity of the three channels, we can represent grayscale in a more general form:

$$v = 0.299r + 0.578g + 0.114b \qquad (3)$$

RDH-GI uses the R and B channels of the colour image to embed information and ensures grayscale invariance by adjusting the pixel value of the G channel [9, 13]. Specifically, we add secret information in the R channel and additional error correction bits in the B channel to distinguish the original information of the pixels to recover the host histogram and solve the problem of grayscale conflicts [13]. When the R and B channels are embedded with information and become $r'$ and $b'$, to ensure that $v$ remains unchanged, we should only adjust the value of the $g$ channel to $g'$ as Eq (4) shows.

$$g' = (v - 0.299r' - 0.114b')/0.587 \qquad (4)$$

When $r'$ and $b'$ are restored to R and B, reuse the formula (4) to calculate the restored G. However, the G obtained at this time is different from the original value of channel G (the reason for rounding [13]), the absolute value of the difference is 0 or 1, so we need to record this difference in the channel B when embedding to prepare for the recovery.

**Prediction Error Generation.** Prediction errors should be calculated for the R and B channels before embedding messages. Predictors such as median-edge detector (MED) [27] or accurate gradient selective prediction (AGSP) [24] are designed to compute the scale of channels R and B, as shown in Fig. 2. We found that the choice of predictor had a limited impact on the final RAEs, so we adopted a simpler approach, which was proposed by [13]. This new MED-liked structure only considers the relationship between the total 4 pixels (AGSP needs to consider 9) and proposes a 3rd-degree polynomial predictor that makes predictions based on the grey values of pixels in each channel in Eq (5) format. This polynomial replaces the pixel comparison operation of MED.

$$P_{MED(i,j)} = \begin{matrix} a \\ b \\ c \end{matrix} \begin{bmatrix} 1, & v_{i,j}, & v_{i,j}^2 \end{bmatrix} \qquad (5)$$

where $i,j$ represents the position of the predicted pixel in the channel. $P_{MED(i,j)}$ is the prediction value based on the grayscale input $v$. Set elements a,b and c are the parameters of adaptive optimization.

To avoid overfitting or underfitting when the predicted value deviates from the real value. The MED-like method proposed by [13] sets the following limits for the predicted value so that it will not deviate too much from the standard, as shown in Eq (6).

$$\hat{r}_{ij} = \begin{cases} min(r_{i+1,j}, r_{i,j+1}) & P_{MED(i,j)} \leq min(r_{i+1,j}, r_{i,j+1}) \\ max(r_{i+1,j}, r_{i,j+1}) & P_{MED(i,j)} \geq max(r_{i+1,j}, r_{i,j+1}) \\ P_{MED(i,j)} & otherwise. \end{cases} \qquad (6)$$

Thus we can get predicted errors and ready for embedding secret data by calculating $(r_{i,j} - \hat{r}_{i,j})$. The authenticated receiver can reverse the above process to obtain target information.

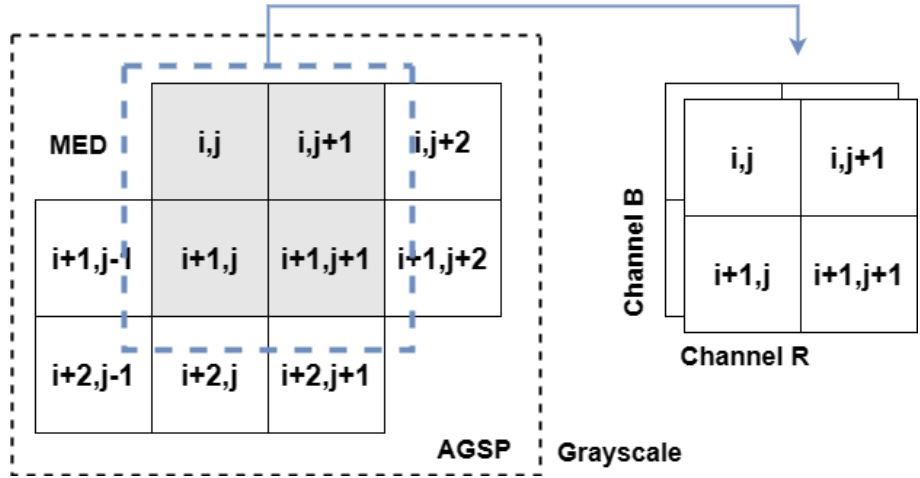

Fig. 2: Workflow of the predictor MGD.

Table 1: Comparison of Queries on ImageNet and CIFAR-10.

| Attack Method | ImageNet | CIFAR-10 |
|---|---|---|
| SimBA-DCT | 8320 | 348 |
| AutoZoom | 13260 | 260 |
| GenAttack | 10421 | 851 |
| HSJA | $> 20K$ | $> 5K$ |
| 3BS-Attack | 7841 | 308 |
| 5BS-Attack | 8234 | 325 |
| 3BS*-Attack | 8105 | 311 |

## 4 Experiments

### 4.1 Datasets and Implementation

Our experiments focus on the targeted attack setting in the black-box scenario. We evaluate the success rates and visual quality of reversible adversarial example (RAE) and adversarial example (AE) on ImageNet [8], and CIFAR-10 [15]. We randomly sample 1000 images from ImageNet and 30,000 from CIFAR-10 as the test set. The victim model is unified as ResNet50 [12] pretrained on CIFAR-10 and ImageNet.

### 4.2 Attack Ability Analysis

**Adversarial Attack.** We test the attack success rate and the average number of queries of the beam search ($n$BS-Attack, $n$ = size of the beam, $*$ means decay coefficient = 1/4, the default is 1/3) attack in table 1. We chose two beam sizes to ensure the query efficiency and selected different decay as the ablation. We also perform an evaluation using various attack methods, including SimBA-DCT [11], GenAttack [1], AutoZoom [25], HopSkipJumpAttack (HSJA) [6]. All frameworks listed above perform targeted black-box attacks.

Table 1 shows the average number of queries required for these methods to achieve a 95 % success rate. The result indicates that our method can effectively limit the number of queries to about 8000 times on ImageNet. The query counts on CIFAR-10 can also be stabilized at around 300. It should be noted that the time for beam search to traverse the output gain is less than the time to query the attack result. The former only increases the linear time complexity based on the original method. Taking CIFAR-10 as an example, an image can reduce the running time by about 11ms (comparing SimBA-DCT and 3BS-Attack), demonstrating that the beam search attack reduces the time complexity while reducing the number of queries.

We demonstrate part of the training process in Fig. 3. The left and right figures show a comparison of attack success rates when the beam size is 3 and 5, respectively. We can see that our method outperforms the SOTA method SimBA in Table 1 on almost all timestamps, especially when the number of queries is small. This indicates that our method has high query efficiency.

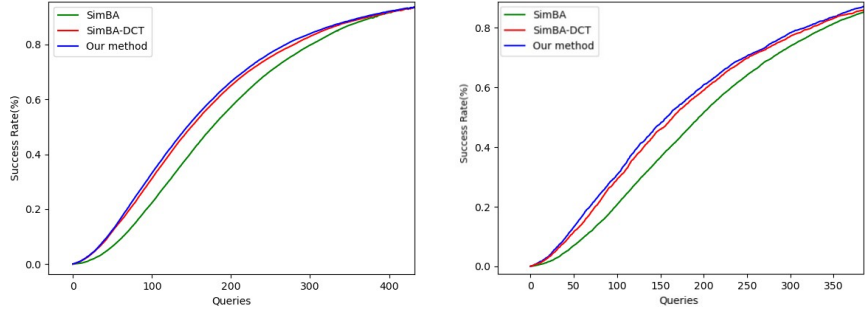

Fig. 3: An example of comparison between our method (3/5BS-Attack) and SimBA/SimBA-DCT [11] against ResNet50.

Table 2: Comparison of success rate (%) for ImageNet.

| Attack Method | white-box - | black-box untargeted | targeted |
|---|---|---|---|
| FGSM-RHM | 0.8034 | 0.4942 | 0.1751 |
| BIM-RHM | 0.9527 | 0.2731 | 0.1536 |
| C&W-RHM | 0.9842 | < 0.05 | − |
| C&W-RIT | 0.9621 | < 0.05 | − |
| Deepfool-RIT | 0.5785 | < 0.01 | − |
| Our Method | 0.9441 | 0.8963 | 0.3321 |

**Reversible Adversarial Attack.** We measured the attack success rate of our RAE framework and compared it with the RIT-based (reversible image transformation) RAE method (RIT-RAE) [30] and the RHM-based (recursive histogram modification) RAE method (RHM-RAE) [19] to evaluate its performance. RIT algorithm [33] inverts the source image into a camouflage image of the same size to achieve privacy. RHM [32] tries to establish the equivalence between reversible data hiding and lossless data compression to improve the final effect of the RDH framework. The above RAE methods combine RDH technology with existing adversarial attack methods, including FGSM [10], BIM [16], C&W [5], and Deefool [20]. All RAE frames used for comparison are baseline with the highest success rate (regardless of the visual effects of the generated RAEs).

Table 2 shows the attack success rates of various RAEs under different settings. FGSM/BIM - RHM seems to have considerable accuracy when performing black-box untargeted attacks because it generates sizeable adversarial perturbations. Hence, the corresponding RAEs have noticeable abnormal pixels. The success rate of other methods in the black-box setting can be nearly negligible. Our method has a higher success rate in the black-box setting. In the extreme case when performing the black-box targeted attack, our RAE can also maintain a success rate of about 30% without sacrificing much of the visual effect of the image. Our method does not involve gradient operations, so we can achieve similar success rates when executing white-box attacks and black-box untargeted attacks, which is impossible with gradient-based methods (such as FGSM-RHM).

### 4.3 Ablation Study

In this part, we mainly perform an ablation study for the proposed beam search attack due to existing RAEs being difficult to implement black-box attacks efficiently. We use 30,000 images to compare the performance of our beam search attack under different parameter settings for CIFAR-10. We have set up our beam search attack with three different parameters: Beam size = 3, beam step = 1/3 step size; beam size = 5, beam step = 1/3 step size; beam size = 3, beam step = 1/4 step size. We test their performance in the black-box setting with targeted attacks against various classifiers, including ResNet-18 and ResNet-50 [12], DenseNet-121 [14] and GoogleNet [23]. We also provide experimental data on SimBA. When BS-Attack removes the beam search structure, the remaining components are similar to SimBA, which can be used for comparison. The experimental results are shown in Table 3 (The symbol '∗' means a 1/4 step size).

It can be seen that all the beam settings in the table reduce the number of model queries while maintaining considerable accuracy. Reducing the beam step will slightly increase the number of queries. Empirically, we suggest that the product of beam size and beam step should be set to about 1, and the training effect of the model is relatively good at this time.

Table 3: Ablation study on CIFAR-10 with different model.

|  | GoogleNet | | ResNet18 | | ResNet50 | | DenseNet121 | |
| --- | --- | --- | --- | --- | --- | --- | --- | --- |
|  | avg.query | success | avg.query | success | avg.query | success | avg.query | success |
| SimBA | 227.05 | 1.0 | 373.10 | 0.999 | 332.34 | 0.998 | 394.29 | 0.999 |
| 3BS | 218.55 | 0.998 | 338.12 | 0.998 | 308.85 | 0.998 | 360.64 | 0.997 |
| 5BS | 223.42 | 0.999 | 347.23 | 0.998 | 311.95 | 0.998 | 375.45 | 0.997 |
| 3BS∗ | 221.72 | 1.0 | 341.38 | 1.0 | 311.58 | 0.998 | 359.23 | 0.998 |

### 4.4 Visual Results

We randomly selected 1000 images on ImageNet to perform the RAE attack with a specific target. Figure 4 shows part adversarial and reversible adversarial examples generated by our method. For each example, we show it in the order of source image (up), adversarial example (middle) and reversible adversarial example (down). Under each picture are the prediction tag and the confidence. As shown in the figure, our RAE method based on RDH-GI hardly changes the visual quality of the image. In the RDH-GI-based framework, all RAE images have the same grayscale version as the adversarial examples (AE). Therefore, when the generated adversarial perturbations cannot be perceived, RAE will be challenging to distinguish from the source image.

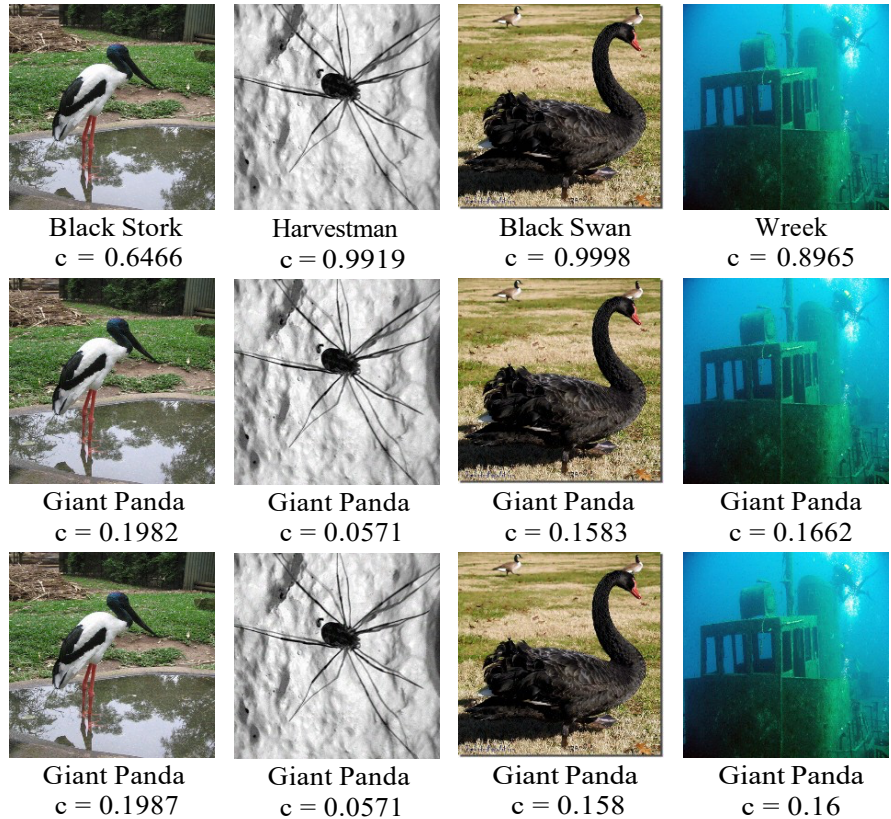

Fig. 4: Examples of original images (up), adversarial examples generated by beam search attack (middle), and our reversible adversarial examples (down).

### 4.5 Visual Quality Analysis

Peak Signal to Noise Ratio (PSNR) is the ratio of peak signal energy to noise average energy. It is a method to measure image quality. To evaluate the visual quality of RAEs, we measure the average PSNR value between RAEs, source images, and adversarial examples (AEs). We also compared the average PSNR values of RIT-RAE [30], and the RHM-RAE [19] under the same setting. To keep consistent with these methods, we embedded about 120000 bits of data in the proposed RAEs. The experimental results are shown in Table 4. Generally, PSNR higher than 40dB indicates that the image is very close to the original image. The value of about 30dB indicates the perceived distortion in the image. It can be seen from Table 4 that the average PSNR between our RAEs and the source image is greater than 40dB, which is higher than other methods listed in Table 4. It is worth noting that most of the current RAEs methods are difficult to execute under the black-box setting. In Table 4, only our methods perform targeted black-box attacks, while the other methods perform white-box attacks. The PSNR between our RAEs and AEs is also considerable, which shows that RDH-GI technology can maintain the visual effect of images while embedding a large amount of information. Therefore, it can be proved that the performance of our RAE is better than most of the existing RAE frameworks, and it can successfully perform targeted attacks in the black-box setting.

Table 4: Comparison of PSNR (dB) between RAEs, AEs and source image.

| method | RAEs/Source | RAEs/AEs |
|---|---|---|
| FGSM-RHM | 29.41 | 34.76 |
| BIM-RHM | 41.19 | 52.83 |
| C&W-RHM | 34.52 | 38.44 |
| C&W-RIT | 37.71 | 35.00 |
| Deepfool-RIT | 35.62 | 35.48 |
| Our Method | 43.15 | 55.30 |

## 5 Conclusion

In this paper, we propose a novel framework to generate reversible adversarial examples with grayscale invariance based on the beam search attack. This framework combines grayscale-invariant reversible data hiding techniques with adversarial attacks, which successfully conduct targeted attacks in a black-box setting for the first time. Our proposed beam search attack exploits the historical information of adversarial perturbations to reduce the number of queries. Experimental results demonstrate that our RAE maintains a high visual similarity to the source image. Our proposed method provides a novel and efficient idea for reversible adversarial attacks in black-box settings. In the future, we intend to improve the method of generating adversarial noise to enhance the performance of reversible adversarial examples.